\documentclass[prl,times,amsmath,amssymb,twocolumn,showpacs,superscriptaddress]{revtex4}
\usepackage[usenames]{color}
\usepackage{amssymb,hyperref}
\usepackage{grffile}
\usepackage[pdftex]{graphicx}
\DeclareGraphicsExtensions{.pdf} 
\usepackage{amsmath, amstext, amssymb, amsfonts, amsxtra}
\usepackage{textcomp}
\usepackage{xspace}
\usepackage{icomma}
\usepackage[short]{datetime}
\usepackage[normalem]{ulem}
\usepackage{pdfpages}

\def \up{{\uparrow}}
\def \down{{\downarrow}}

\def \ell{{d}}

\newcommand{\sign}[1]{~e^{i{\boldsymbol\pi}\cdot #1}~}
\def \av#1{{\langle#1\rangle}}

\newcommand{\bra}[1]{\ensuremath{\langle #1 \vert}\xspace}%

\newcommand{\ket}[1]{\ensuremath{\vert #1 \rangle}\xspace}%
\newcommand{\aver}[1]{\ensuremath{\langle #1 \rangle}\xspace}%

\newcommand{\nop}{\hat{n}}

\newcommand*{\bfrac}[2]{\genfrac{}{}{0pt}{1}{#1}{#2}}

\def \rdi{ {{\bf r}+{\bf d}}}
\def \di{{\bf d}}

\def \r{ {\bf{r}} }

\def \hole{ {0} }
\def \dub{ {\down\up} }

\newcommand{\unige}{D\'epartement de Physique Th\'eorique, Universit\'e de Gen\`eve, CH-1211 Gen\`eve, Switzerland.}
\newcommand{\ubc}{Department of Physics and Astronomy, University of British Columbia, Vancouver V6T 1Z1, Canada.} 

\newcommand{\sutd}{Singapore University of Technology and Design, 20 Dover Drive, 138682 Singapore.}

\begin{document}

\title{Emergence of long distance pair coherence through incoherent\\
local environmental coupling}
\author{Jean-S\'ebastien Bernier}
\affiliation{\ubc} 
\author{Peter Barmettler}
\affiliation{\unige}
\author{Dario Poletti}
\affiliation{\sutd}
\author{Corinna Kollath}
\affiliation{\unige}

\begin{abstract}
We demonstrate that quantum coherence can be generated by the
interplay of coupling to an incoherent environment and kinetic processes.
This joint effect even occurs in a repulsively interacting fermionic system 
initially prepared in an incoherent Mott insulating state. In this case, 
coupling a dissipative noise field to the local spin density produces coherent 
pairs of fermions. The generated pair coherence extends over long distances as 
typically seen in Bose-Einstein condensates. This conceptually surprising
approach provides a novel path towards a better control of
quantum many-body correlations.
\end{abstract}



\maketitle


In recent years, various experimental methods have been developed to 
dynamically generate non-trivial correlations in quantum materials. 
On the one hand, external electromagnetic fields have been used 
to photo-induce phase transitions in solid state materials~\cite{BasovHaule2011}. 
For example, spin density wave order was induced in the normal state of a pnictide compound 
using femtosecond optical pulses~\cite{KimLeitenstorfer2012}. 
A Josephson plasmon, typically present in a superconducting state, has even been 
triggered in a non-superconducting striped-order cuprate by the application of 
mid-infrared femtosecond pulses~\cite{FaustiCavalleri2010}. On the other hand, 
environmental tailoring \cite{MuellerZoller2012} has been used to prepare highly entangled 
states such as a Bell state of two ions~\cite{BarreiroBlatt2011} or
a Tonks-Giradeau-like state in a molecular quantum gas~\cite{BauerLettner2008}. 
In these examples, the realization of complex states relies on the same principle 
as optical pumping whereby atoms are prepared in so-called dark states immune 
to environmental coupling.

\begin{figure}[t]
\includegraphics[width=0.95\columnwidth]{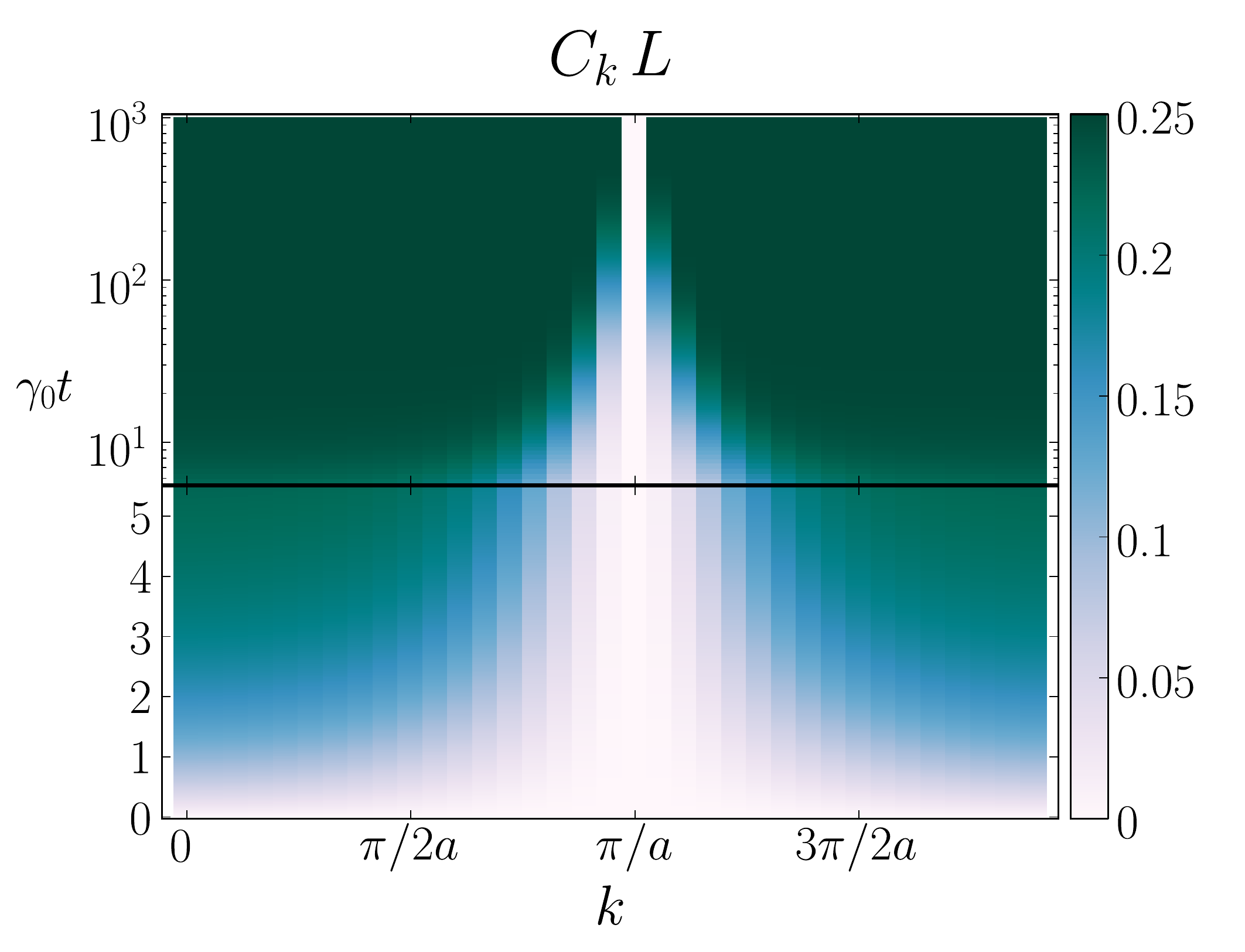}
\caption{Time evolution of the momentum distribution of local pairs. A chain of $36$ lattice 
sites is prepared at $t=0$ in a perfect Mott insulating state where pair correlations are absent. 
A fast build-up in occupation of the momenta except close to $k=\frac{\pi}{a}$ takes place
at short times (plotted on a linear time scale). Then, over time, all momenta except $k=\frac{\pi}{a}$ become 
homogeneously occupied, signaling the generation of the coherence of pairs over longer distances 
(plotted on a logarithmic scale). The evolution is obtained using the effective diffusion equation 
\eqref{eq:diff} with $\frac{U}{\hbar\Gamma}=1.5$ and $\gamma_0=\frac{8J^2}{\hbar^2\Gamma}$.
\label{fig:Fig1}} 
\end{figure}

We report on a complementary mechanism where dynamical generation of 
coherence is achieved through the combined effect of a simple local dissipative 
coupling and kinetic processes. To examplify the inner workings of this mechanism,
we consider an ultracold fermionic gas in an optical lattice subjected to local spin-polarization 
measurements carried out by phase-contrast imaging~\cite{ShinKetterle2006} or spatial and temporal 
light field fluctuations. The dissipative coupling heats up the system 
and destroys single-particle correlations. At the same time the number of local  pairs, which 
are immune to the dissipative coupling, increase due to kinetic hopping. 
Unexpectedly, these local pairs then act as a source for the generation of pair 
correlations over longer distances. The produced correlations are long lived and reminiscent 
to those of the celebrated $\eta$-pairing state~\cite{Yang1989,RoschVojta2008}, a condensate of
bound on-site pairs of momentum $k=\frac{\pi}{a}$ (with $a$ the lattice spacing).
Moreover, the appearance of a sharp feature in the pair momentum distribution, 
as shown in Fig.~\ref{fig:Fig1}, serves as a signature for the formation of long distance coherence. 
In cold atom experiments, such pair momentum distributions can be observed by the projection 
of the local pairs onto molecules~\cite{RegalJin2004, ZwierleinKetterle2004}. As our proposal 
relies both on dissipation and kinetic processes, it is conceptually very different from previous approaches where 
the $\eta$-pairing state was stabilized through either adiabatic state 
preparation~\cite{RoschVojta2008, KantianZoller2010}, or the imprint of phase coherence 
between neighboring sites by a tailored environment~\cite{DiehlZoller2008, MuellerZoller2012}. 


The system under consideration here is made of repulsively interacting fermions 
on a $d$-dimensional lattice of volume $V$ and lattice constant $a$.
We describe this many-body system by the Hubbard model
\begin{eqnarray} 
H \!\! &=& \!\! -J\!\!\!\sum_{\langle {\bf r},\,{\bf r}'\rangle,\sigma} 
      \!\left(\hat c_{{\bf r},\sigma}^\dagger \hat c^{\phantom{\dagger}}_{{\bf r}',\sigma}+\mbox{h.c.}\right)
      + U\!\sum _{\bf r} \hat{n}_{{\bf r},\uparrow} \hat{n}_{{\bf r},\downarrow}\,, \nonumber
\end{eqnarray}
where $\hat c_{\r,\sigma}^\dagger$ is the creation operator for a fermion with spin
$\sigma=\uparrow\!,\downarrow$ and site index $\r$, $\hat{n}_{\r,\sigma} = \hat c_{\r,\sigma}^\dagger
\hat c^{\phantom{\dagger}}_{\r,\sigma}$ is the density operator, $J>0$ is the hopping coefficient, $U$ the 
interaction strength, and $\langle {\bf r},\,{\bf r}'\rangle$ indicates that the sum is done over 
nearest-neighbors. This Hamiltonian is one of the simplest models capturing the interplay 
between the kinetic and interaction energies, and can be used, for example, to understand
the metal to Mott insulator transition. A particularly clean realization 
of this model is achieved using ultracold fermionic gases confined to optical 
lattices~\cite{KoehlEsslinger2005}.

In the present work, we assume the system to be initially prepared in a stationary state of this 
Hamiltonian, typically a Mott insulator as realized in Refs.~\cite{SchneiderRosch2008, JordensEsslinger2008}. 
We study the system dynamics after coupling a dissipative environment to the local spin densities such that
\begin{equation}
\frac{d}{dt}\hat{\rho}(t)=-\frac i \hbar \big[\hat{H},\,\hat{\rho}(t)\big]
+ \mathcal{D}\left[\hat{\rho}\left(t\right)\right]\,, \label{eq:master}     
\end{equation}
with 
\begin{equation} 
\mathcal{D} \left[\hat{\rho}(t)\right]
=\Gamma \sum_\r \left(\nop_{s,\r}\hat{\rho}~\nop_{s,\r}-\frac 1 2 \nop_{s,\r}^2~\hat{\rho}
-\frac 1 2 \hat{\rho}~\nop_{s,\r}^2\right)\,.\label{eq:dissipator}
\end{equation}
The quantum jump operators $\nop_{s,\r}=\hat{n}_{\uparrow,\r}-\hat{n}_{\downarrow,\r}$ 
measure the local spin polarization. The dissipative coupling $\mathcal{D}$ can be realized, for example, by a 
light field whose frequency is chosen in between the transitions of the two fermionic 
states as used for phase contrast imaging~\cite{ShinKetterle2006}. 
This light field can either be used to probe the spin density locally, e.g. in combination with an independent 
addressing of each site~\cite{BakrGreiner2009,Sherson2010} or to create ``magnetic field'' noise, e.g. through the 
realization of spatially-disordered and time-decorrelated white noise patterns. 

This dissipative mechanism
leads to an exponential decay of single particle correlations as
$ \aver{\hat c^\dagger_{\r,\sigma}\hat c_{\rdi,\sigma}}\propto e^{-\Gamma t}$. In contrast, pair correlations
$C_\di= \frac{1}{V}\sum_\r\av{\hat c_{\r\down}^\dagger\hat c_{\r\up}^\dagger\hat c_{\rdi\up}\hat c_{\rdi\down}}$,
remain unchanged under the action of the dissipator, $\mathcal{D}$, as 
doublons (doubly occupied sites) and holes (empty sites), which have no net polarization, 
belong to the dissipation-free subspace.
In particular, the $\eta$-paired state, generated through the repeated application of the operator 
$\hat \eta^\dagger = \sum_\r \sign{\r} \hat{c}_{\r\uparrow}^\dagger \hat{c}_{\r\downarrow}^\dagger$ 
on the vacuum, is part of this subspace; here ${{\boldsymbol \pi}} = (\pi, \dots, \pi)$.
As the $\eta$-paired state is an eigenstate of the Hamiltonian, it is
immune against the action of both the unitary and dissipative operators~(see Eq.~\eqref{eq:master}). 
Interestingly, even if the initial state does not overlap with the $\eta$-paired state,
correlations may emerge due to kinetic processes. 


\begin{figure}[th]
\includegraphics[width=0.95\columnwidth]{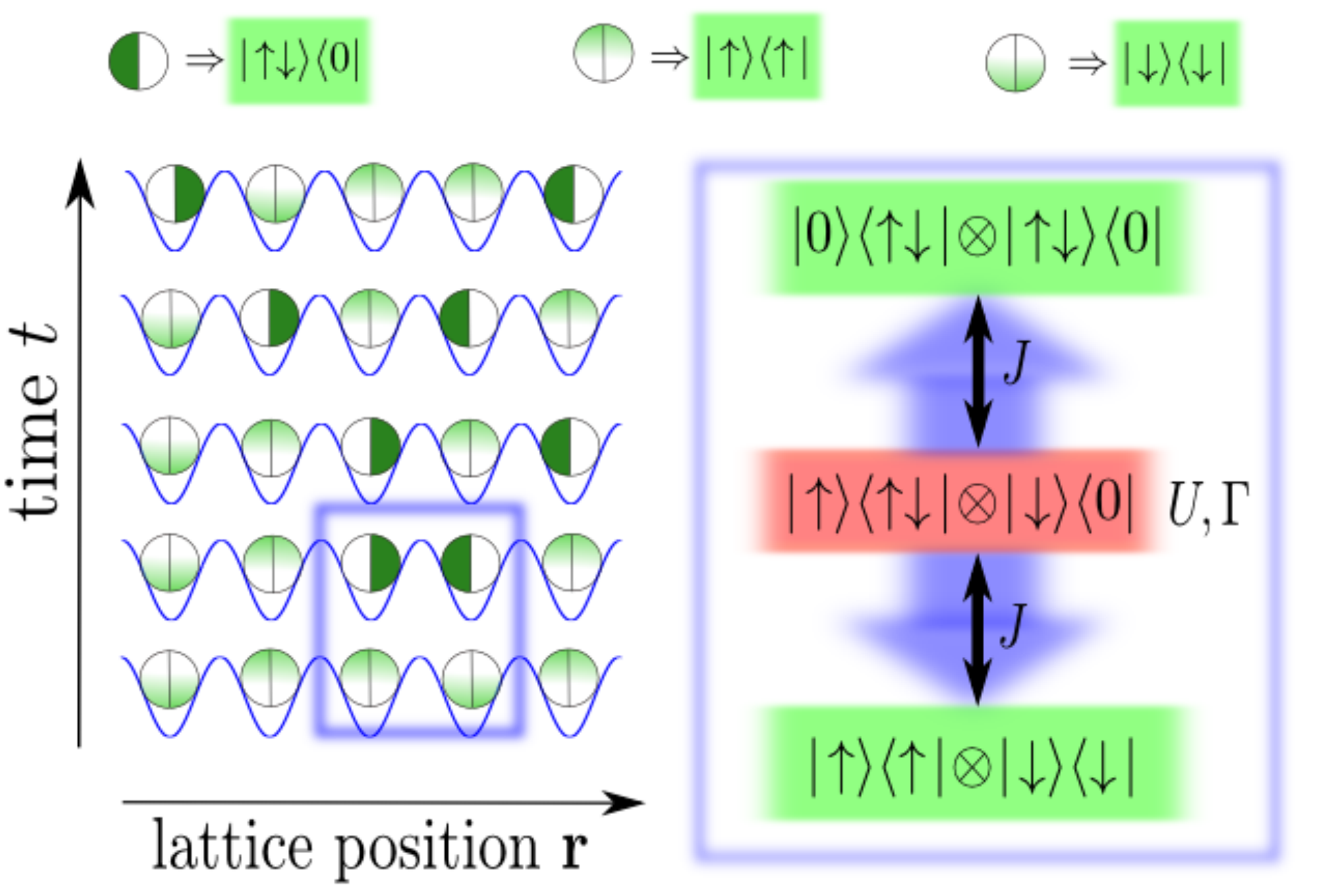}
\caption{Left: Example of the effective creation and diffusion of pair correlations. 
Right: Within the adiabatic elimination method, the evolution is based on the effective coupling of two states (lower and upper state) 
of the decoherence free subspace via a virtual excitation (center). The virtual state is reached by 
the hopping process and can decay with $\Gamma$ and dephase due to interaction $U$. Here this process
is examplified for a state with no pair correlations (lower state) connected to a state containing pair 
correlations (upper state) through the creation process (box) presented on the left panel.  
\label{fig:Sketch}} 
\end{figure}

Irrespective of the coupling strength and the properties of the Hamiltonian, at sufficiently 
large times, $\Gamma t \gg 1$, the dissipation free subspace is reached. This subspace is 
highly degenerate with respect to the dissipator $\mathcal{D}$ and the Hamiltonian can lift this degeneracy.
To understand the dynamics, we
perform adiabatic elimination (see Appendix) 
revealing how hopping-induced virtual excitations, 
around the dissipation-free subspace, affect the evolution of the system (cf.~Fig.~\ref{fig:Sketch}).
 The effective coupling via the virtual excitations depends on whether the interaction 
energy is changed during the process and takes the form
\begin{equation}
\gamma_0 = \frac{8{J^2}}{\hbar^2 \Gamma}\;\;\;\text{and}\;\;\;\gamma_U=  \frac{8J^2\Gamma}{\hbar^2\Gamma^2+U^2}\,. 
\nonumber
\end{equation}

Using this perturbative approach, 
the equations describing 
the evolution of staggered pair correlations, i.e.~$\tilde C_\di=-\sign{\di} C_\di + \frac{\delta_{\di,0}}{4}$, 
for times larger than $\frac{1}{\Gamma}$ are cast into a system of coupled diffusion equations (cf.~Fig.~\ref{fig:Sketch}, left panel):
\begin{eqnarray}
\frac{d}{dt}\tilde C^{}_{\di}(t) = 
\!\!\!\!\!\!\sum_{{\di'},\,{|\di-\di'|=1}}\!\!\!\! 
A_{\di',\di}(t)\left(\tilde C^{}_{\di'}(t) -\tilde C^{}_{\di}(t)\right).
\label{eq:diff}
\end{eqnarray}
The diffusion constant depends on the coupling to the different virtual excitations weighted by their 
probability to occur,
\begin{eqnarray}
A_{\di',\di}(t) \equiv D(t) =  
\gamma_0\!\left(\frac{1}{4}+\tilde C_0(t)\right)\!+\gamma_{U}\!\left(\frac{1}{4}-\tilde C_0(t)\!\right)
\nonumber
\end{eqnarray}
for $|\di|$ and $|\di'| \neq 0$, while
$A_{\di',{\bf 0}}(t)=A_{{\bf 0},\di}(t) = \frac{\gamma_U}{2}$.
For the sake of concreteness, we assumed above that the system was half filled and 
translationally invariant. 
Generalizations are straightforward and do not lead to qualitative changes. 
Moreover, we provide numerical evidence that this diffusive behavior is not restricted 
to the domain $\Gamma \gg \frac{J}{\hbar}$, but is valid even at weak coupling $\Gamma < \frac{J}{\hbar}$. 
However, in the latter case, the diffusion constant deviates from the 
perturbative results.

\begin{figure}[!ht]
\includegraphics[width=0.95\columnwidth]{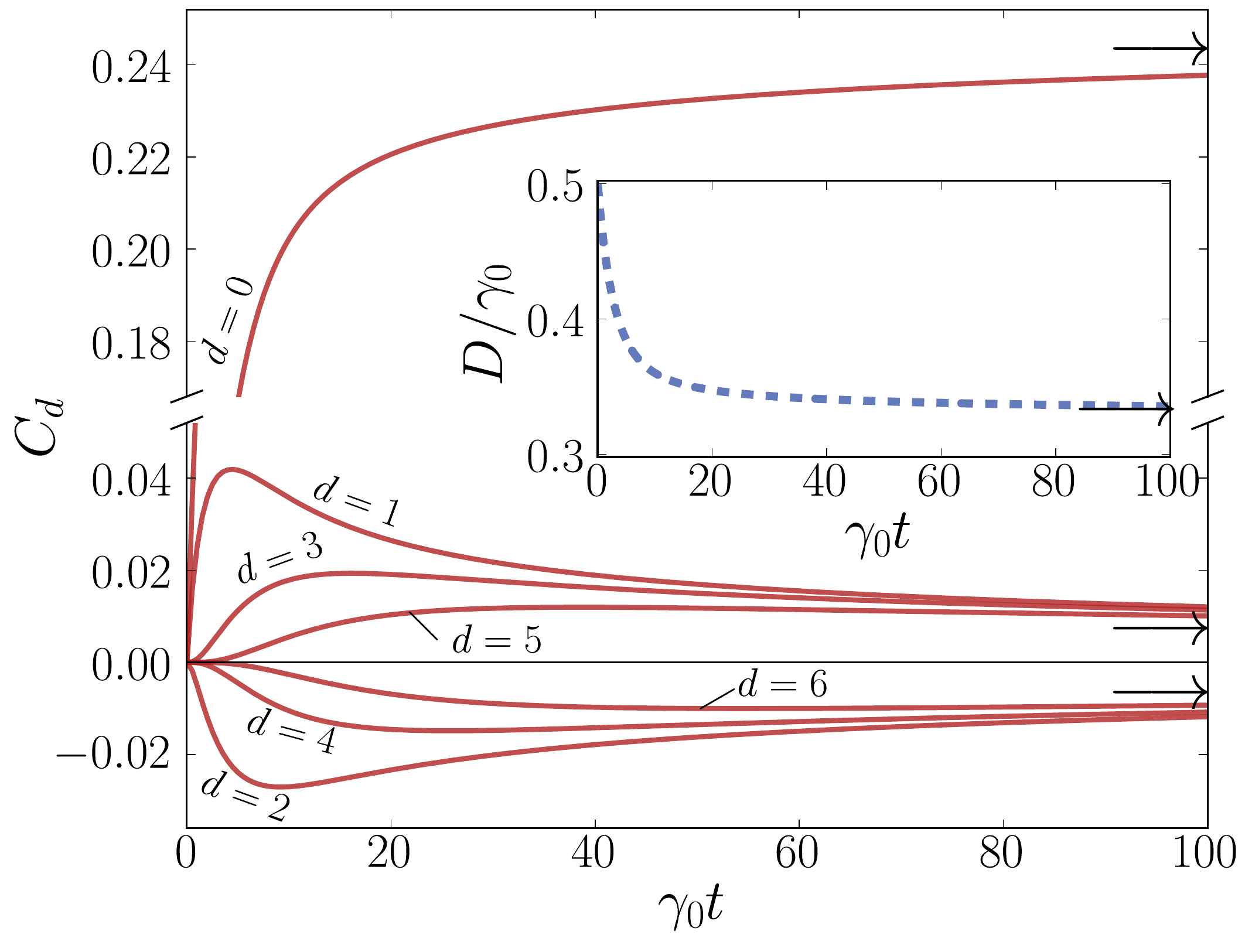}
\caption{$C_d(t)$ as a function of time $t$ as described by the 
diffusion equation \eqref{eq:diff}. The chain is prepared in a perfect Mott insulator with 
$L=36$ sites and evolves with $\frac{U}{\hbar\Gamma}=1.5$. Arrows mark the $t\rightarrow\infty$ limit. 
The double occupancy, $C_{d=0}$, and nearest-neighbor pair 
correlation, $C_{d=1}$, raise quickly and feed the delayed increase of correlations at large 
distances. Inset: evolution of the diffusion constant as a function of time.
$D(t)$ becomes time-independent as the double occupancy, $C_{d=0}$, saturates.
\label{fig:diffusion_vs_t}}      
\end{figure}

\begin{figure}[t]
\includegraphics[width=0.95\columnwidth]{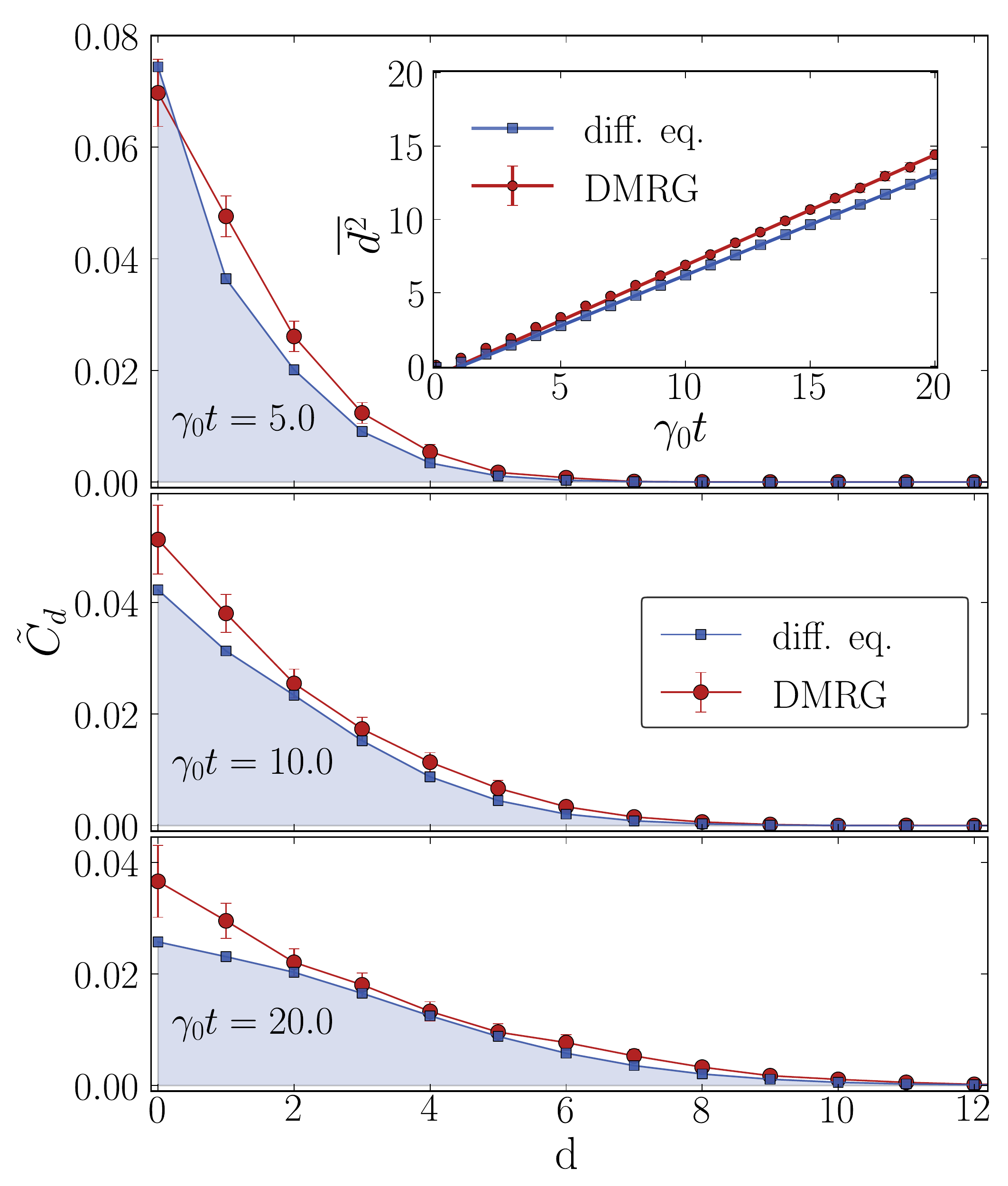}
\caption{ $\tilde{C}_d(t)$, the staggered pair correlations, 
are shown as a function of distance, $d$, for three different times in a chain of $L=36$. 
We compare the solution of the diffusion equation \eqref{eq:diff} with time-dependent DMRG simulations. 
For both cases the initial conditions are taken from ground state DMRG calculations at $U=12J$. 
We use $\frac{U}{\hbar\Gamma}=1.5$ and in the time-dependent DMRG $U=12J$. 
Inset: Symbols represent the variance of the pair correlation distribution versus time.  
Lines are linear fits for $10<\gamma_0 t<20$.
\label{fig:diffusion_vs_d}} 
\end{figure}

We illustrate the creation of correlations using, as an example, a system initially 
in a Mott insulating state, i.e. $C_\di(0) = 0$ for all $\di$. 
In Fig.~\ref{fig:diffusion_vs_t} we depict the dynamics 
triggered by the action of the Hamiltonian, $\hat H$, and the dissipator, $\mathcal{D}$, on a
one-dimensional system. First, double occupancy and short range pair correlations rise on the time 
scale $\frac{1}{\gamma_0}$.
Then, following this 
initial build-up, the double occupancy and the nearest-neighbor pair correlation 
act as sources for the propagation of pair correlations over longer distances. 
Within the perturbatively derived Eq.~\eqref{eq:diff}, one expects the propagation of the 
staggered pair correlations to be described by a normal diffusion process. 
The evolution of the system is constrained by the constant of motion
$F=C_{{\bf k}=\pi/a}$, where $C_{{\bf k}}=\frac{1}{V} \sum_\di ~e^{-ia{\bf k}\cdot\di}~ C_\di$ is the 
momentum distribution of the local pairs (Fig.~\ref{fig:Fig1}). The constant of motion is 
proportional to the number of $\eta$-pairs, $\langle \hat \eta^\dagger \hat\eta \rangle$, initially 
present in the system.
For an atomic Mott insulator, $\langle\hat \eta^\dagger \hat \eta \rangle = 0$.
Thus, the pair correlations asymptotic values 
are $C_\di(t\rightarrow\infty) = -\sign{\di}\frac{1}{4V}$ 
for $|\di| \neq 0$ while $C_0(t\rightarrow\infty)=\frac{1}{4}\left(1-\frac{1}{V}\right)$. 
The ``overall'' sign of $C_{\di\neq 0}$, and thus the phase slip between $d=0$ and $d=1$ depends on the initial state and would differ if 
$F > \frac{1}{4}$.
We note that the saturation of the double occupancy implies that 
the diffusion constant becomes time-independent: 
$D(t\rightarrow \infty) \sim \frac{\gamma_0+\gamma_U}{4}$ (see inset in Fig.~\ref{fig:diffusion_vs_t}).
To study this diffusive spreading, 
we plot in Fig.~\ref{fig:diffusion_vs_d} 
the staggered correlations at different times versus distance. 
In terms of $\tilde C_\di$, the initial Mott insulator is characterized by a peak at $\di=0$. 
As shown on Fig.~\ref{fig:diffusion_vs_d}, the interplay of dissipation and hopping gradually 
transforms this peak into a broad gaussian distribution. 

To check the validity of our 
perturbative results against unbiased methods, we use time-dependent DMRG to solve the master 
equation \eqref{eq:master} stochastically
~\footnote{The timestep used in the 
integration via Suzuki-Trotter decomposition is about $0.02 \frac{1}{\Gamma}$. 
Numerical accuracy was ensured by retaining between $400$ DMRG states (for $\frac{\hbar\Gamma}{J}=8$) 
and $2500$ states (for $\frac{\hbar\Gamma}{J}=0.25$) and we typically sampled over a few 
thousand stochastic realizations. }.
One can appreciate in Fig.~\ref{fig:diffusion_vs_d} that
the pair correlations obtained from DMRG are
in good agreement with the pertubative results.
Deviations away from the expected gaussian distribution mostly occur at 
short distances. This discrepancy is partially attributed to the decoupling of density 
correlations applied in the derivation of Eq.~\eqref{eq:diff}.

The diffusive propagation is best characterized by the variance 
\begin{eqnarray}
\overline{\di^2(t)} = \sum_\di \di^2\tilde C_\di(t)/\sum_\di\tilde C_\di(0)\,.
\nonumber
\end{eqnarray}
Generally, for a hypercubic lattice with connectivity $z$, the variance of a diffusive process obeys 
\begin{eqnarray}
\frac{d}{dt}\overline{\di^2(t)} = z D(t) \,.
\nonumber
\end{eqnarray}
Within the perturbative treatment, at $t \gg \frac{1}{\gamma_0}$ 
where $D(t)$ becomes constant, the variance should rise linearly with time. 
The variance, shown in the inset of Fig.~\ref{fig:diffusion_vs_d} confirms 
this statement. Small deviations from the linear behavior are consistent with the 
time-dependence of $D(t)$.

\begin{figure}[t]
\includegraphics[width=0.98\columnwidth]{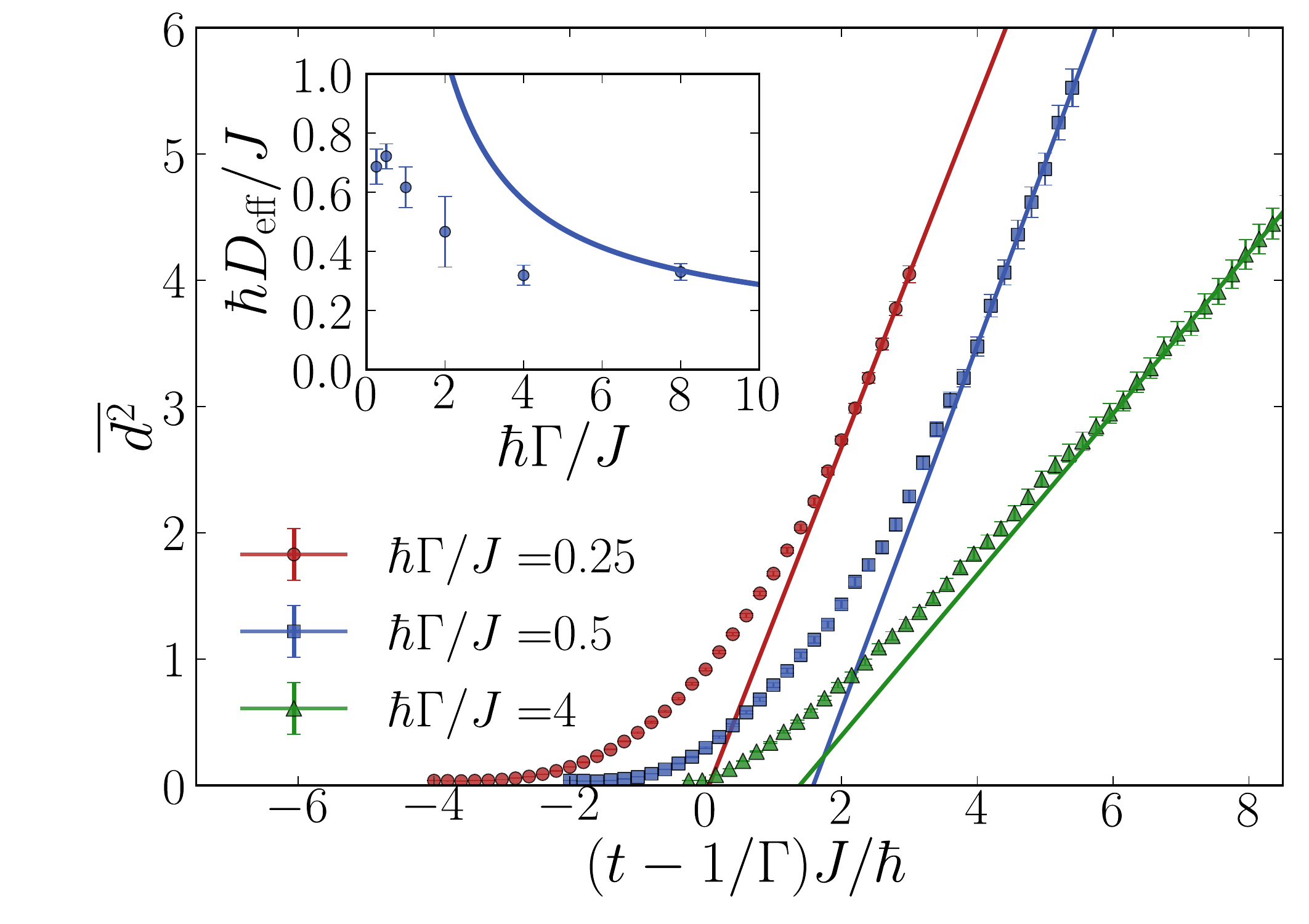}
\caption{The variance of the distribution of the staggered pair correlations $\tilde{C}_d(t)$ with distance for different 
dissipative strength $\frac{\hbar\Gamma}{J}$ at constant $U=12J$ in a chain of length $36$. 
The linear rise of the variance at larger times is in accordance with a diffusive formation of the pair 
correlations. Symbols are taken from DMRG simulations, straight lines are linear fits 
$\overline{d^2}= 2 D_{\mbox{\scriptsize eff}}\,t + \mbox{const.}$ for values with $\overline{d^2}>2.5 $. 
Inset: $D_{\mbox{\scriptsize eff}}$ as a function of dissipative coupling at $U=12J$ \label{fig:d2}(symbols). 
The straight line corresponds to $D(t\rightarrow\infty)$ derived within perturbation theory and agrees 
well with the numerical results for large coupling strengths.}      
\end{figure}

Remarkably, we find the diffusive
description of the propagation of pair correlations to remain 
valid down to the weakly dissipative regime. 
However, in this regime, the diffusion constants need to be phenomenologically
determined. In Fig.~\ref{fig:d2}, we study a strongly interacting system $U=12J$ with couplings 
from $\Gamma = 4 \frac{J}{\hbar}$ down to $\Gamma = 0.25 \frac{J}{\hbar}$. In all cases, a normal diffusive regime is entered after 
$t \sim \frac{1}{\Gamma}$.  As shown in the inset of Fig.~\ref{fig:d2}, the effective diffusion constants, 
in the strong dissipative coupling limit, agree nicely with our perturbative predictions. 
As expected, with decreasing $\Gamma$, the deviation from the analytic predictions increases.
Nevertheless, the correct qualitative behavior is predicted for $\Gamma\gtrsim\frac{J}{\hbar}$: 
the effective diffusion constant increases with $\frac{1}{\Gamma}$. For  $\Gamma\lesssim\frac{J}{\hbar}$ 
our simulations suggest a saturation of the diffusion constant to roughly $0.7\frac{J}{\hbar}$.

In summary, we demonstrated here that the combined action of incoherent local environmental 
coupling and kinetic processes can result in the emergence of long distance pair correlations 
in repulsive fermionic lattice systems. In contrast to correlations realized in cooled 
condensed matter systems, typically sensitive to temperature, 
the non-equilibrium mechanism presented above is immune against thermal fluctuations. 
This conceptually surprising approach provides a new route towards a better control of
quantum many-body correlations.


We thank E. Demler, T. Esslinger, V. Galitski, A. Kantian, M. K\"ohl, H. Pichler, A. Rosch and 
P. Zoller for fruitful discussions, and A. Georges for his valuable contributions to the early 
part of this work. We acknowledge ANR (FAMOUS), SNSF under Division II, MaNEP, CIFAR and 
NSERC for their financial support and the KITP (grant no.~NSF PHY11-25915) 
for its hospitality. 


\appendix*
\section{Appendix: Adiabatic elimination}

Locally, the dissipation free subspace with respect to $\mathcal{D}$ 
is spanned by the diagonal operators 
$\{\hat p_{\r,\hole}\,=\,\ket{0}\bra{0}\,,\,\hat p_{\r,\down}\,=\,\ket{\down} \bra{\down}\,,\, 
\hat p_{\r,\up}\,=\,\ket{\up} \bra{\up}\,,\,\hat p_{\r,\up\down}\,=\,\ket{\up\down} \bra{\up\down}\}$, 
and the off-diagonal operators $\{\hat d_\r\,=\,\ket{0} \bra{\up\down}\,,\,\hat d_\r^\dag 
\,=\,\ket{\up\down} \bra{0}\}$ annihilating 
or creating a pair at site $\r$ such that 
$\hat{d}_\r^\dagger =  \hat{c}_{\r\up}^\dagger \hat{c}_{\r\down}^\dagger$. In addition, the
first excited subpace can be defined via the basis elements 
$\{\hat p_{\r,\hole}\hat c_{\r,\sigma}\,,\,\hat c_{\r,\sigma}^\dagger \hat p_{\r,\hole}\,,\,
\hat p_{\r,\dub}\hat c_{\r,\sigma}^\dagger\,,\,\hat c_{\r,\sigma}\hat p_{\r,\up\down}\}$. 
These operators form a diagonal basis for $-\frac{iU}{ \hbar }\big[ \hat{p}_{\r,\up\down},\, . \big]
+ \mathcal{D} \left[.\right]$  with eigenvalues
$\lambda_\alpha\in\{-\frac{\Gamma}{2},-\frac{\Gamma}{2},-\frac{\Gamma}{2}+i\frac{U}{\hbar},
-\frac{\Gamma}{2}-i\frac{U}{\hbar}\}$, $\alpha=1,~\dots,~4$. 
We associate a projector $\mathcal{P}^\alpha_\r$ to each subspace associated with a certain eigenvalue of the 
excited subspace, while $\mathcal{P}^0_\r$ projects onto the dissipation free subspace.  
Via adiabatic elimination of the excited subspace (see e.g. \cite{Garcia-RipollCirac2009}), 
one derives effective equations of motion for the basis elements within the dissipation free subspace, for example
\begin{eqnarray}
\frac{d}{dt}\hat d_\r &=& 
\!\!\!\!\!\!
\sum_{\bfrac{\r',\,|\r-\r'|=1}{\alpha,\alpha'=1,\dots,4}}
\!\!
\frac{J^2}{\hbar^2}\frac{\mathcal{P}^0_\r\mathcal{P}^0_{\r'}
\Big[\hat K_{\r,\r'}\,,\,\mathcal{P}^{\alpha}_{\r}\mathcal{P}^{\alpha'}_{\r'}
\Big[\hat K_{\r,\r'}\,,\,\hat d_\r\Big]\Big]}{\lambda_{\alpha}+\lambda_{\alpha'}+i\frac{U}{\hbar}} \nonumber \\
&=& \sum_{\r',\,|\r-\r'|=1}\hat A_{\r'}\hat{d}_\r+\hat A_{\r}\hat{d}_{\r'}\,, \label{eq:leffd}
\end{eqnarray}
\begin{eqnarray}
\text{with}\;\;\hat A_\r=-2\frac{J^2}{\hbar^2}
\left\{\frac{(\hat{p}_{\r,\up}+\hat p_{\r,\down})}{\Gamma} + \frac{\hat{p}_{\r,\dub}}{\Gamma + i\frac{U}{\hbar}} 
+\frac{\hat{p}_{\r,\hole}}{\Gamma- i\frac{U}{\hbar}}\right\} \nonumber
\end{eqnarray}
and $\hat K_{\r,\r'}=\hat c_{{\bf r},\sigma}^\dagger \hat c^{\phantom{\dagger}}_{{\bf r}',\sigma}+\mbox{h.c.}$
Eq.~\eqref{eq:leffd} is used to derive the equation of motion 
for $\hat d_\r^\dagger\hat d_{\r'}$ with $|\r - \r'| > 1$. A similar procedure is
used to find the equations for $|\r-\r'| \leq 1$. In order to construct the closed set of equations of 
motion for the pair correlators, Eq.~\eqref{eq:diff}, we decouple $\av{\hat p_{\r,n} \hat O_{\r'}}$ 
as $\av{\hat p_{\r,n}}\av{\hat O_{\r'}}$ where
$n = \{0,\,\up,\,\down,\,\up\down \}$ and $\hat O_{\r'}$ is an arbitrary local operator 
on site $\r'\neq\r$.


\begin{thebibliography}{19}%
\makeatletter
\providecommand \@ifxundefined [1]{%
 \@ifx{#1\undefined}
}%
\providecommand \@ifnum [1]{%
 \ifnum #1\expandafter \@firstoftwo
 \else \expandafter \@secondoftwo
 \fi
}%
\providecommand \@ifx [1]{%
 \ifx #1\expandafter \@firstoftwo
 \else \expandafter \@secondoftwo
 \fi
}%
\providecommand \natexlab [1]{#1}%
\providecommand \enquote  [1]{``#1''}%
\providecommand \bibnamefont  [1]{#1}%
\providecommand \bibfnamefont [1]{#1}%
\providecommand \citenamefont [1]{#1}%
\providecommand \href@noop [0]{\@secondoftwo}%
\providecommand \href [0]{\begingroup \@sanitize@url \@href}%
\providecommand \@href[1]{\@@startlink{#1}\@@href}%
\providecommand \@@href[1]{\endgroup#1\@@endlink}%
\providecommand \@sanitize@url [0]{\catcode `\\12\catcode `\$12\catcode
  `\&12\catcode `\#12\catcode `\^12\catcode `\_12\catcode `\%12\relax}%
\providecommand \@@startlink[1]{}%
\providecommand \@@endlink[0]{}%
\providecommand \url  [0]{\begingroup\@sanitize@url \@url }%
\providecommand \@url [1]{\endgroup\@href {#1}{\urlprefix }}%
\providecommand \urlprefix  [0]{URL }%
\providecommand \Eprint [0]{\href }%
\@ifxundefined \urlstyle {%
  \providecommand \doi  [0]{\begingroup \@sanitize@url \@doi}%
  \providecommand \@doi [1]{\endgroup \@@startlink {\doibase
  #1}doi:\discretionary {}{}{}#1\@@endlink }%
}{%
  \providecommand \doi  [0]{doi:\discretionary{}{}{}\begingroup
  \urlstyle{rm}\Url }%
}%
\providecommand \doibase [0]{http://dx.doi.org/}%
\providecommand \Doi [0]{\begingroup \@sanitize@url \@Doi }%
\providecommand \@Doi  [1]{\endgroup\@@startlink{\doibase#1}\@@Doi}%
\providecommand \@@Doi [1]{#1\@@endlink}%
\providecommand \selectlanguage [0]{\@gobble}%
\providecommand \bibinfo  [0]{\@secondoftwo}%
\providecommand \bibfield  [0]{\@secondoftwo}%
\providecommand \translation [1]{[#1]}%
\providecommand \BibitemOpen [0]{}%
\providecommand \bibitemStop [0]{}%
\providecommand \bibitemNoStop [0]{.\EOS\space}%
\providecommand \EOS [0]{\spacefactor3000\relax}%
\providecommand \BibitemShut  [1]{\csname bibitem#1\endcsname}%
\bibitem [{\citenamefont {Basov}\ \emph {et~al.}(2011)\citenamefont {Basov},
  \citenamefont {Averitt}, \citenamefont {van~der Marel}, \citenamefont
  {Dressel},\ and\ \citenamefont {Haule}}]{BasovHaule2011}%
  \BibitemOpen
  \bibfield  {author} {\bibinfo {author} {\bibfnamefont {D.~N.}\ \bibnamefont
  {Basov}}, \bibinfo {author} {\bibfnamefont {R.~D.}\ \bibnamefont {Averitt}},
  \bibinfo {author} {\bibfnamefont {D.}~\bibnamefont {van~der Marel}}, \bibinfo
  {author} {\bibfnamefont {M.}~\bibnamefont {Dressel}}, \ and\ \bibinfo
  {author} {\bibfnamefont {K.}~\bibnamefont {Haule}},\ }\Doi
  {10.1103/RevModPhys.83.471} {\bibfield  {journal} {\bibinfo  {journal} {Rev.
  Mod. Phys.},\ }\textbf {\bibinfo {volume} {83}},\ \bibinfo {pages} {471}
  (\bibinfo {year} {2011})}\BibitemShut {NoStop}%
\bibitem [{\citenamefont {Kim}\ \emph {et~al.}(2012)\citenamefont {Kim},
  \citenamefont {Pashkin}, \citenamefont {Sch\"afer}, \citenamefont {Beyer},
  \citenamefont {Porer}, \citenamefont {Wolf}, \citenamefont {Bernhard},
  \citenamefont {Demsar}, \citenamefont {Huber},\ and\ \citenamefont
  {Leitenstorfer}}]{KimLeitenstorfer2012}%
  \BibitemOpen
  \bibfield  {author} {\bibinfo {author} {\bibfnamefont {K.~W.}\ \bibnamefont
  {Kim}}, \bibinfo {author} {\bibfnamefont {A.}~\bibnamefont {Pashkin}},
  \bibinfo {author} {\bibfnamefont {H.}~\bibnamefont {Sch\"afer}}, \bibinfo
  {author} {\bibfnamefont {M.}~\bibnamefont {Beyer}}, \bibinfo {author}
  {\bibfnamefont {M.}~\bibnamefont {Porer}}, \bibinfo {author} {\bibfnamefont
  {T.}~\bibnamefont {Wolf}}, \bibinfo {author} {\bibfnamefont {C.}~\bibnamefont
  {Bernhard}}, \bibinfo {author} {\bibfnamefont {J.}~\bibnamefont {Demsar}},
  \bibinfo {author} {\bibfnamefont {R.}~\bibnamefont {Huber}}, \ and\ \bibinfo
  {author} {\bibfnamefont {A.}~\bibnamefont {Leitenstorfer}},\ }\href@noop {}
  {\bibfield  {journal} {\bibinfo  {journal} {Nature Materials},\ }\textbf
  {\bibinfo {volume} {11}},\ \bibinfo {pages} {497} (\bibinfo {year}
  {2012})}\BibitemShut {NoStop}%
\bibitem [{\citenamefont {Fausti}\ \emph {et~al.}(2011)\citenamefont {Fausti},
  \citenamefont {Tobey}, \citenamefont {Dean}, \citenamefont {Kaiser},
  \citenamefont {Dienst}, \citenamefont {Hoffmann}, \citenamefont {Pyon},
  \citenamefont {Takayama}, \citenamefont {Takagi},\ and\ \citenamefont
  {Cavalleri}}]{FaustiCavalleri2010}%
  \BibitemOpen
  \bibfield  {author} {\bibinfo {author} {\bibfnamefont {D.}~\bibnamefont
  {Fausti}}, \bibinfo {author} {\bibfnamefont {R.~I.}\ \bibnamefont {Tobey}},
  \bibinfo {author} {\bibfnamefont {N.}~\bibnamefont {Dean}}, \bibinfo {author}
  {\bibfnamefont {S.}~\bibnamefont {Kaiser}}, \bibinfo {author} {\bibfnamefont
  {A.}~\bibnamefont {Dienst}}, \bibinfo {author} {\bibfnamefont {M.~C.}\
  \bibnamefont {Hoffmann}}, \bibinfo {author} {\bibfnamefont {S.}~\bibnamefont
  {Pyon}}, \bibinfo {author} {\bibfnamefont {T.}~\bibnamefont {Takayama}},
  \bibinfo {author} {\bibfnamefont {H.}~\bibnamefont {Takagi}}, \ and\ \bibinfo
  {author} {\bibfnamefont {A.}~\bibnamefont {Cavalleri}},\ }\Doi
  {10.1126/science.1197294} {\bibfield  {journal} {\bibinfo  {journal}
  {Science},\ }\textbf {\bibinfo {volume} {331}},\ \bibinfo {pages} {189}
  (\bibinfo {year} {2011})}\BibitemShut {NoStop}%
\bibitem [{\citenamefont {M\"uller}\ \emph {et~al.}(2012)\citenamefont
  {M\"uller}, \citenamefont {Diehl}, \citenamefont {Pupillo},\ and\
  \citenamefont {Zoller}}]{MuellerZoller2012}%
  \BibitemOpen
  \bibfield  {author} {\bibinfo {author} {\bibfnamefont {M.}~\bibnamefont
  {M\"uller}}, \bibinfo {author} {\bibfnamefont {S.}~\bibnamefont {Diehl}},
  \bibinfo {author} {\bibfnamefont {G.}~\bibnamefont {Pupillo}}, \ and\
  \bibinfo {author} {\bibfnamefont {P.}~\bibnamefont {Zoller}},\ }\href@noop {}
  {\bibfield  {journal} {\bibinfo  {journal} {Advances in Atomic, Molecular,
  and Optical Physics},\ }\textbf {\bibinfo {volume} {61}},\ \bibinfo {pages}
  {1} (\bibinfo {year} {2012})}\BibitemShut {NoStop}%
\bibitem [{\citenamefont {Barreiro}\ \emph {et~al.}(2011)\citenamefont
  {Barreiro}, \citenamefont {M{\"u}ller}, \citenamefont {Schindler},
  \citenamefont {Nigg}, \citenamefont {Monz}, \citenamefont {Chwalla},
  \citenamefont {Hennrich}, \citenamefont {Roos}, \citenamefont {Zoller},\ and\
  \citenamefont {Blatt}}]{BarreiroBlatt2011}%
  \BibitemOpen
  \bibfield  {author} {\bibinfo {author} {\bibfnamefont {J.}~\bibnamefont
  {Barreiro}}, \bibinfo {author} {\bibfnamefont {M.}~\bibnamefont
  {M{\"u}ller}}, \bibinfo {author} {\bibfnamefont {P.}~\bibnamefont
  {Schindler}}, \bibinfo {author} {\bibfnamefont {D.}~\bibnamefont {Nigg}},
  \bibinfo {author} {\bibfnamefont {T.}~\bibnamefont {Monz}}, \bibinfo {author}
  {\bibfnamefont {M.}~\bibnamefont {Chwalla}}, \bibinfo {author} {\bibfnamefont
  {M.}~\bibnamefont {Hennrich}}, \bibinfo {author} {\bibfnamefont
  {C.}~\bibnamefont {Roos}}, \bibinfo {author} {\bibfnamefont {P.}~\bibnamefont
  {Zoller}}, \ and\ \bibinfo {author} {\bibfnamefont {R.}~\bibnamefont
  {Blatt}},\ }\href@noop {} {\bibfield  {journal} {\bibinfo  {journal}
  {Nature},\ }\textbf {\bibinfo {volume} {470}},\ \bibinfo {pages} {486}
  (\bibinfo {year} {2011})}\BibitemShut {NoStop}%
\bibitem [{\citenamefont {Bauer}\ \emph {et~al.}(2008)\citenamefont {Bauer},
  \citenamefont {Syassen}, \citenamefont {Dietze}, \citenamefont {Volz},
  \citenamefont {Rempe}, \citenamefont {Garcia-Ripoll}, \citenamefont {Cirac},
  \citenamefont {D\"urr},\ and\ \citenamefont {Lettner}}]{BauerLettner2008}%
  \BibitemOpen
  \bibfield  {author} {\bibinfo {author} {\bibfnamefont {D.~M.}\ \bibnamefont
  {Bauer}}, \bibinfo {author} {\bibfnamefont {N.}~\bibnamefont {Syassen}},
  \bibinfo {author} {\bibfnamefont {D.}~\bibnamefont {Dietze}}, \bibinfo
  {author} {\bibfnamefont {T.}~\bibnamefont {Volz}}, \bibinfo {author}
  {\bibfnamefont {G.}~\bibnamefont {Rempe}}, \bibinfo {author} {\bibfnamefont
  {J.-J.}\ \bibnamefont {Garcia-Ripoll}}, \bibinfo {author} {\bibfnamefont
  {J.~I.}\ \bibnamefont {Cirac}}, \bibinfo {author} {\bibfnamefont
  {S.}~\bibnamefont {D\"urr}}, \ and\ \bibinfo {author} {\bibfnamefont
  {M.}~\bibnamefont {Lettner}},\ }\enquote {\bibinfo {title} {A dissipative
  Tonks-Girardeau gas of molecules},}\ in\ \Doi {10.1142/9789814273008_0029}
  {\emph {\bibinfo {booktitle} {Pushing the frontiers of atomic physics}}}\
  (\bibinfo  {publisher} {World Scientific, Singapore},\ \bibinfo {year}
  {2008})\ Chap.~\bibinfo {chapter} {29}, pp.\ \bibinfo {pages}
  {307--314}\BibitemShut {NoStop}%
\bibitem [{\citenamefont {Shin}\ \emph {et~al.}(2006)\citenamefont {Shin},
  \citenamefont {Zwierlein}, \citenamefont {Schunck}, \citenamefont
  {Schirotzek},\ and\ \citenamefont {Ketterle}}]{ShinKetterle2006}%
  \BibitemOpen
  \bibfield  {author} {\bibinfo {author} {\bibfnamefont {Y.}~\bibnamefont
  {Shin}}, \bibinfo {author} {\bibfnamefont {M.~W.}\ \bibnamefont {Zwierlein}},
  \bibinfo {author} {\bibfnamefont {C.~H.}\ \bibnamefont {Schunck}}, \bibinfo
  {author} {\bibfnamefont {A.}~\bibnamefont {Schirotzek}}, \ and\ \bibinfo
  {author} {\bibfnamefont {W.}~\bibnamefont {Ketterle}},\ }\Doi
  {10.1103/PhysRevLett.97.030401} {\bibfield  {journal} {\bibinfo  {journal}
  {Phys. Rev. Lett.},\ }\textbf {\bibinfo {volume} {97}},\ \bibinfo {pages}
  {030401} (\bibinfo {year} {2006})}\BibitemShut {NoStop}%
\bibitem [{\citenamefont {Yang}(1989)}]{Yang1989}%
  \BibitemOpen
  \bibfield  {author} {\bibinfo {author} {\bibfnamefont {C.~N.}\ \bibnamefont
  {Yang}},\ }\Doi {10.1103/PhysRevLett.63.2144} {\bibfield  {journal} {\bibinfo
   {journal} {Phys. Rev. Lett.},\ }\textbf {\bibinfo {volume} {63}},\ \bibinfo
  {pages} {2144} (\bibinfo {year} {1989})}\BibitemShut {NoStop}%
\bibitem [{\citenamefont {Rosch}\ \emph {et~al.}(2008)\citenamefont {Rosch},
  \citenamefont {Rasch}, \citenamefont {Binz},\ and\ \citenamefont
  {Vojta}}]{RoschVojta2008}%
  \BibitemOpen
  \bibfield  {author} {\bibinfo {author} {\bibfnamefont {A.}~\bibnamefont
  {Rosch}}, \bibinfo {author} {\bibfnamefont {D.}~\bibnamefont {Rasch}},
  \bibinfo {author} {\bibfnamefont {B.}~\bibnamefont {Binz}}, \ and\ \bibinfo
  {author} {\bibfnamefont {M.}~\bibnamefont {Vojta}},\ }\Doi
  {10.1103/PhysRevLett.101.265301} {\bibfield  {journal} {\bibinfo  {journal}
  {Phys. Rev. Lett.},\ }\textbf {\bibinfo {volume} {101}},\ \bibinfo {pages}
  {265301} (\bibinfo {year} {2008})}\BibitemShut {NoStop}%
\bibitem [{\citenamefont {Regal}\ \emph {et~al.}(2004)\citenamefont {Regal},
  \citenamefont {Greiner},\ and\ \citenamefont {Jin}}]{RegalJin2004}%
  \BibitemOpen
  \bibfield  {author} {\bibinfo {author} {\bibfnamefont {C.~A.}\ \bibnamefont
  {Regal}}, \bibinfo {author} {\bibfnamefont {M.}~\bibnamefont {Greiner}}, \
  and\ \bibinfo {author} {\bibfnamefont {D.~S.}\ \bibnamefont {Jin}},\ }\Doi
  {10.1103/PhysRevLett.92.040403} {\bibfield  {journal} {\bibinfo  {journal}
  {Phys. Rev. Lett.},\ }\textbf {\bibinfo {volume} {92}},\ \bibinfo {pages}
  {040403} (\bibinfo {year} {2004})}\BibitemShut {NoStop}%
\bibitem [{\citenamefont {Zwierlein}\ \emph {et~al.}(2004)\citenamefont
  {Zwierlein}, \citenamefont {Stan}, \citenamefont {Schunck}, \citenamefont
  {Raupach}, \citenamefont {Kerman},\ and\ \citenamefont
  {Ketterle}}]{ZwierleinKetterle2004}%
  \BibitemOpen
  \bibfield  {author} {\bibinfo {author} {\bibfnamefont {M.~W.}\ \bibnamefont
  {Zwierlein}}, \bibinfo {author} {\bibfnamefont {C.~A.}\ \bibnamefont {Stan}},
  \bibinfo {author} {\bibfnamefont {C.~H.}\ \bibnamefont {Schunck}}, \bibinfo
  {author} {\bibfnamefont {S.~M.~F.}\ \bibnamefont {Raupach}}, \bibinfo
  {author} {\bibfnamefont {A.~J.}\ \bibnamefont {Kerman}}, \ and\ \bibinfo
  {author} {\bibfnamefont {W.}~\bibnamefont {Ketterle}},\ }\Doi
  {10.1103/PhysRevLett.92.120403} {\bibfield  {journal} {\bibinfo  {journal}
  {Phys. Rev. Lett.},\ }\textbf {\bibinfo {volume} {92}},\ \bibinfo {pages}
  {120403} (\bibinfo {year} {2004})}\BibitemShut {NoStop}%
\bibitem [{\citenamefont {Kantian}\ \emph {et~al.}(2010)\citenamefont
  {Kantian}, \citenamefont {Daley},\ and\ \citenamefont
  {Zoller}}]{KantianZoller2010}%
  \BibitemOpen
  \bibfield  {author} {\bibinfo {author} {\bibfnamefont {A.}~\bibnamefont
  {Kantian}}, \bibinfo {author} {\bibfnamefont {A.~J.}\ \bibnamefont {Daley}},
  \ and\ \bibinfo {author} {\bibfnamefont {P.}~\bibnamefont {Zoller}},\ }\Doi
  {10.1103/PhysRevLett.104.240406} {\bibfield  {journal} {\bibinfo  {journal}
  {Phys. Rev. Lett.},\ }\textbf {\bibinfo {volume} {104}},\ \bibinfo {pages}
  {240406} (\bibinfo {year} {2010})}\BibitemShut {NoStop}%
\bibitem [{\citenamefont {Diehl}\ \emph {et~al.}(2008)\citenamefont {Diehl},
  \citenamefont {Micheli}, \citenamefont {Kantian}, \citenamefont {Kraus},
  \citenamefont {B{\"u}chler},\ and\ \citenamefont {Zoller}}]{DiehlZoller2008}%
  \BibitemOpen
  \bibfield  {author} {\bibinfo {author} {\bibfnamefont {S.}~\bibnamefont
  {Diehl}}, \bibinfo {author} {\bibfnamefont {A.}~\bibnamefont {Micheli}},
  \bibinfo {author} {\bibfnamefont {A.}~\bibnamefont {Kantian}}, \bibinfo
  {author} {\bibfnamefont {B.}~\bibnamefont {Kraus}}, \bibinfo {author}
  {\bibfnamefont {H.}~\bibnamefont {B{\"u}chler}}, \ and\ \bibinfo {author}
  {\bibfnamefont {P.}~\bibnamefont {Zoller}},\ }\href@noop {} {\bibfield
  {journal} {\bibinfo  {journal} {Nature Physics},\ }\textbf {\bibinfo {volume}
  {4}},\ \bibinfo {pages} {878} (\bibinfo {year} {2008})}\BibitemShut {NoStop}%
\bibitem [{\citenamefont {K\"ohl}\ \emph {et~al.}(2005)\citenamefont {K\"ohl},
  \citenamefont {Moritz}, \citenamefont {St\"oferle}, \citenamefont
  {G\"unter},\ and\ \citenamefont {Esslinger}}]{KoehlEsslinger2005}%
  \BibitemOpen
  \bibfield  {author} {\bibinfo {author} {\bibfnamefont {M.}~\bibnamefont
  {K\"ohl}}, \bibinfo {author} {\bibfnamefont {H.}~\bibnamefont {Moritz}},
  \bibinfo {author} {\bibfnamefont {T.}~\bibnamefont {St\"oferle}}, \bibinfo
  {author} {\bibfnamefont {K.}~\bibnamefont {G\"unter}}, \ and\ \bibinfo
  {author} {\bibfnamefont {T.}~\bibnamefont {Esslinger}},\ }\href@noop {}
  {\bibfield  {journal} {\bibinfo  {journal} {Phys.~ Rev.~ Lett.},\ }\textbf
  {\bibinfo {volume} {94}},\ \bibinfo {pages} {080403} (\bibinfo {year}
  {2005})}\BibitemShut {NoStop}%
\bibitem [{\citenamefont {Schneider}\ \emph {et~al.}(2008)\citenamefont
  {Schneider}, \citenamefont {Hackerm{\"u}ller}, \citenamefont {Will},
  \citenamefont {Best}, \citenamefont {Bloch}, \citenamefont {Costi},
  \citenamefont {Helmes}, \citenamefont {Rasch},\ and\ \citenamefont
  {Rosch}}]{SchneiderRosch2008}%
  \BibitemOpen
  \bibfield  {author} {\bibinfo {author} {\bibfnamefont {U.}~\bibnamefont
  {Schneider}}, \bibinfo {author} {\bibfnamefont {L.}~\bibnamefont
  {Hackerm{\"u}ller}}, \bibinfo {author} {\bibfnamefont {S.}~\bibnamefont
  {Will}}, \bibinfo {author} {\bibfnamefont {T.}~\bibnamefont {Best}}, \bibinfo
  {author} {\bibfnamefont {I.}~\bibnamefont {Bloch}}, \bibinfo {author}
  {\bibfnamefont {T.~A.}\ \bibnamefont {Costi}}, \bibinfo {author}
  {\bibfnamefont {R.~W.}\ \bibnamefont {Helmes}}, \bibinfo {author}
  {\bibfnamefont {D.}~\bibnamefont {Rasch}}, \ and\ \bibinfo {author}
  {\bibfnamefont {A.}~\bibnamefont {Rosch}},\ }\href@noop {} {\bibfield
  {journal} {\bibinfo  {journal} {Science},\ }\textbf {\bibinfo {volume}
  {322}},\ \bibinfo {pages} {1520} (\bibinfo {year} {2008})}\BibitemShut
  {NoStop}%
\bibitem [{\citenamefont {J{\"o}rdens}\ \emph {et~al.}(2008)\citenamefont
  {J{\"o}rdens}, \citenamefont {Strohmaier}, \citenamefont {G{\"u}nter},
  \citenamefont {Moritz},\ and\ \citenamefont
  {Esslinger}}]{JordensEsslinger2008}%
  \BibitemOpen
  \bibfield  {author} {\bibinfo {author} {\bibfnamefont {R.}~\bibnamefont
  {J{\"o}rdens}}, \bibinfo {author} {\bibfnamefont {N.}~\bibnamefont
  {Strohmaier}}, \bibinfo {author} {\bibfnamefont {K.}~\bibnamefont
  {G{\"u}nter}}, \bibinfo {author} {\bibfnamefont {H.}~\bibnamefont {Moritz}},
  \ and\ \bibinfo {author} {\bibfnamefont {T.}~\bibnamefont {Esslinger}},\
  }\href@noop {} {\bibfield  {journal} {\bibinfo  {journal} {Nature},\ }\textbf
  {\bibinfo {volume} {455}},\ \bibinfo {pages} {204} (\bibinfo {year}
  {2008})}\BibitemShut {NoStop}%
\bibitem [{\citenamefont {Bakr}\ \emph {et~al.}(2009)\citenamefont {Bakr},
  \citenamefont {Gillen}, \citenamefont {Peng}, \citenamefont {F\"{o}lling},\
  and\ \citenamefont {Greiner}}]{BakrGreiner2009}%
  \BibitemOpen
  \bibfield  {author} {\bibinfo {author} {\bibfnamefont {W.~S.}\ \bibnamefont
  {Bakr}}, \bibinfo {author} {\bibfnamefont {J.~I.}\ \bibnamefont {Gillen}},
  \bibinfo {author} {\bibfnamefont {A.}~\bibnamefont {Peng}}, \bibinfo {author}
  {\bibfnamefont {S.}~\bibnamefont {F\"{o}lling}}, \ and\ \bibinfo {author}
  {\bibfnamefont {M.}~\bibnamefont {Greiner}},\ }\Doi {10.1038/nature08482}
  {\bibfield  {journal} {\bibinfo  {journal} {Nature},\ }\textbf {\bibinfo
  {volume} {462}},\ \bibinfo {pages} {74} (\bibinfo {year} {2009})},\ ISSN
  \bibinfo {issn} {1476-4687}\BibitemShut {NoStop}%
\bibitem [{\citenamefont {Sherson}\ \emph {et~al.}(2010)\citenamefont
  {Sherson}, \citenamefont {Weitenberg}, \citenamefont {Endres}, \citenamefont
  {Cheneau}, \citenamefont {Bloch},\ and\ \citenamefont {Kuhr}}]{Sherson2010}%
  \BibitemOpen
  \bibfield  {author} {\bibinfo {author} {\bibfnamefont {J.~F.}\ \bibnamefont
  {Sherson}}, \bibinfo {author} {\bibfnamefont {C.}~\bibnamefont {Weitenberg}},
  \bibinfo {author} {\bibfnamefont {M.}~\bibnamefont {Endres}}, \bibinfo
  {author} {\bibfnamefont {M.}~\bibnamefont {Cheneau}}, \bibinfo {author}
  {\bibfnamefont {I.}~\bibnamefont {Bloch}}, \ and\ \bibinfo {author}
  {\bibfnamefont {S.}~\bibnamefont {Kuhr}},\ }\Doi {10.1038/nature09378}
  {\bibfield  {journal} {\bibinfo  {journal} {Nature},\ }\textbf {\bibinfo
  {volume} {467}},\ \bibinfo {pages} {68} (\bibinfo {year} {2010})},\ ISSN
  \bibinfo {issn} {1476-4687}\BibitemShut {NoStop}%
\bibitem [{\citenamefont {Garc{\'i}a-Ripoll}\ \emph {et~al.}(2009)\citenamefont
  {Garc{\'i}a-Ripoll}, \citenamefont {D{\"u}rr}, \citenamefont {Syassen},
  \citenamefont {Bauer}, \citenamefont {Lettner}, \citenamefont {Rempe},\ and\
  \citenamefont {Cirac}}]{Garcia-RipollCirac2009}%
  \BibitemOpen
  \bibfield  {author} {\bibinfo {author} {\bibfnamefont {J.~J.}\ \bibnamefont
  {Garc{\'i}a-Ripoll}}, \bibinfo {author} {\bibfnamefont {S.}~\bibnamefont
  {D{\"u}rr}}, \bibinfo {author} {\bibfnamefont {N.}~\bibnamefont {Syassen}},
  \bibinfo {author} {\bibfnamefont {D.~M.}\ \bibnamefont {Bauer}}, \bibinfo
  {author} {\bibfnamefont {M.}~\bibnamefont {Lettner}}, \bibinfo {author}
  {\bibfnamefont {G.}~\bibnamefont {Rempe}}, \ and\ \bibinfo {author}
  {\bibfnamefont {J.~I.}\ \bibnamefont {Cirac}},\ }\href@noop {} {\bibfield
  {journal} {\bibinfo  {journal} {New Journal of Physics},\ }\textbf {\bibinfo
  {volume} {11}},\ \bibinfo {pages} {013053} (\bibinfo {year}
  {2009})}\BibitemShut {NoStop}%
\end{thebibliography}

%

\end{document}